 \documentclass[preprint, superscriptaddress,preprintnumbers,amsmath,amssymb]{revtex4}

\allowdisplaybreaks
\allowdisplaybreaks[4]
\usepackage{graphicx}
\usepackage{dcolumn}
\usepackage{bm}
\usepackage{color}
\usepackage{nicefrac}
\usepackage{upgreek}
\usepackage[dvipdfm,
            pdfstartview=FitH,
            CJKbookmarks=true,
            bookmarksnumbered=true,
            bookmarksopen=true,
            colorlinks=true,
            pdfborder=001,
            linkcolor=blue,
            anchorcolor=blue,
            citecolor=blue
            ]{hyperref}

\begin{document}

\title{Three-level mixing model for nuclear chiral rotation:
Role of planar component}

\author{Q. B. Chen}
\email{qbchen@pku.edu.cn}
\affiliation{State Key Laboratory of Nuclear Physics and Technology,
             School of Physics, Peking University, Beijing 100871, China}%
\affiliation{Physik-Department, Technische Universit\"{a}t
             M\"{u}nchen, D-85747 Garching, Germany}

\author{K. Starosta}
\email{starosta@sfu.ca}
\affiliation{Department of Chemistry, Simon Fraser University,
8888 University Drive, Burnaby, BC, Canada, V5A 1S6}

\author{T. Koike}
\email{tkoike@lambda.phys.tohoku.ac.jp}
\affiliation{Department of Physics, Tohoku University,
6-3 Aoba Aramaki Aoba Sendai, Japan, 980-8578}

\date{\today}

\begin{abstract}

Three- and two-level mixing models are proposed to
  understand the doubling of states at the same spin and parity in
  triaxially-deformed atomic nuclei with odd numbers of protons and
  neutrons.  The Particle-Rotor Model for such nuclei is solved using
  the newly proposed basis which couples angular momenta of two
  valence nucleons and the rotating triaxial mean-field into
  left-handed $|\mathcal{L}\rangle$, right-handed
  $|\mathcal{R}\rangle$, and planar $|\mathcal{P}\rangle$
  configurations.  The presence and the impact of the planar component
  is investigated as a function of the total spin for mass A$\approx$130
  nuclei with the valence h$_{11/2}$ proton particle, valence
  h$_{11/2}$ neutron hole and the maximum difference between principle
  axes allowed by the quadrupole deformation of the mean field.  It is
  concluded that at each spin value the higher-energy member of a
  doublet of states is built on the anti-symmetric combination of
  $|\mathcal{L}\rangle$ and $|\mathcal{R}\rangle$ and is free of the
  $|\mathcal{P}\rangle$ component, indicating that it is of pure
  chiral geometry.  For the lower-energy member of the doublet, the
  contribution of the $|\mathcal{P}\rangle$ component to the
  eigenfunction first decreases and then
  increases as a function of the total spin. This trend as well as the
  energy splitting between the doublet states are both determined by
  the Hamiltonian matrix elements between the planar
  ($|\mathcal{P}\rangle$) and non-planar ($|\mathcal{L}\rangle$ and
  $|\mathcal{R}\rangle$) subspaces of the full Hilbert space.

\end{abstract}


\maketitle

A well known example of a two-level quantum system is the parity
doublet for a particle in a symmetric Double Square Potential Well
(DSPW)~\cite{Sakurai1994book, Feyman1965book}. The Hamiltonian for a
particle in the DSPW commutes with the parity operator $P$. Thus
eigenstates of the Hamiltonian are eigenstates of the parity.  When
the potential barrier between the wells forming the DSPW is finite,
the positive parity ground state $|+\rangle$ has a partner that is of the
negative parity $|-\rangle$ separated by an energy interval
related to the height of the potential barrier. When the barrier
separating the DSPW wells become infinite the $|+\rangle$ and the
$|-\rangle$ partner states become degenerate. In this case linear
combinations of the $|\pm\rangle$ eigenstates can be formed,
\begin{equation}
|\mathcal{L}/\mathcal{R}\rangle=\nicefrac{1}{\sqrt{2}} \left(|+\rangle
\mp |-\rangle\right),
\end{equation}
with the $|\mathcal{L}/\mathcal{R}\rangle$ states being the
eigenstates of the Hamiltonian but not the parity operator. The
$|\mathcal{L}/\mathcal{R}\rangle$ states represent in this particular
case the lowest-energy solution in the infinite potential well on the
right or the left of the DSPW for which
$P|\mathcal{L}\rangle=|\mathcal{R}\rangle$ or vice versa.  In the case
of the infinite barrier the $|\pm\rangle$ as well as the
$|\mathcal{L}/\mathcal{R}\rangle$ states describe system equally well.
Thus, there is a freedom of choice in selecting the former set
of well defined parity, or the latter set for which the parity is not
defined, but the interpretation of the wave function of a
particle confined to either the left or the right well is intuitive.

For the last two decades, a novel two-level quantum mechanical
system involving three axial vectors of angular momenta has been
under intensive scrutiny in relation to coupling of collective and
single particle motions in rotating nuclei~\cite{Frauendorf1997NPA}.
For an odd-odd rotating nucleus with a triaxial shape (a shape of a
kiwi fruit) and valence proton/neutron in high-$j$ particle/hole
orbitals, the total angular momentum $\vec{I}$ may lie outside of
the three principal planes. Consequently, the three components of
the total angular momentum along the principal axes can be oriented
in either left- or right-handed systems, thus defining chirality.
The left- and the right-handed systems are transformed into each
other by the chiral operator which combines spatial rotation by
$180^\circ$ around the intermediate axis with the time reversal,
$\mathcal{R}_2(\pi)\mathcal{T}$~\cite{Frauendorf2001RMP}. The
formation of chiral systems takes place in the reference frame which
is defined by the principle axes of the deformed nuclear mean field.
This is a rotating, thus non-inertial reference frame. In the
stationary, thus inertial, laboratory reference frame, the wave
functions are linear combinations which include with equal
probability the left-handed and right-handed components. This leads
to the doubling of states in the laboratory frame referred to as
``restoration of chiral symmetry''. Consequently, the transformation
to the laboratory frame gives rise to a pair of nearly degenerate
$\Delta I=1$ bands with the same parity, often referred to as chiral
doublet bands~\cite{Frauendorf1997NPA}.

The purpose of this Rapid Communication is to analyze models for the
doubling of states related to chiral coupling of angular momenta
vectors drawing analogies to the DSPW model of parity doublets. The
models presented here, while simplified, capture the essential
physics of nuclear chiral rotation. In particular, the mechanism
leading to energy degeneracy and the role of non-chiral (planar)
component are illustrated. On the other hand, while the behaviour of
the chiral system bears strong analogies to the DSPW, there is one
prominent difference should be emphasized. In the DSPW model the
potential barrier is static, while in the case of nuclear chirality
the barrier height for systems of opposite chirality depends on the
total angular momentum of the rotating nucleus, although it is
static at a given $I$.

\begin{figure}[t]
  \begin{center}
    \includegraphics[width=9 cm]{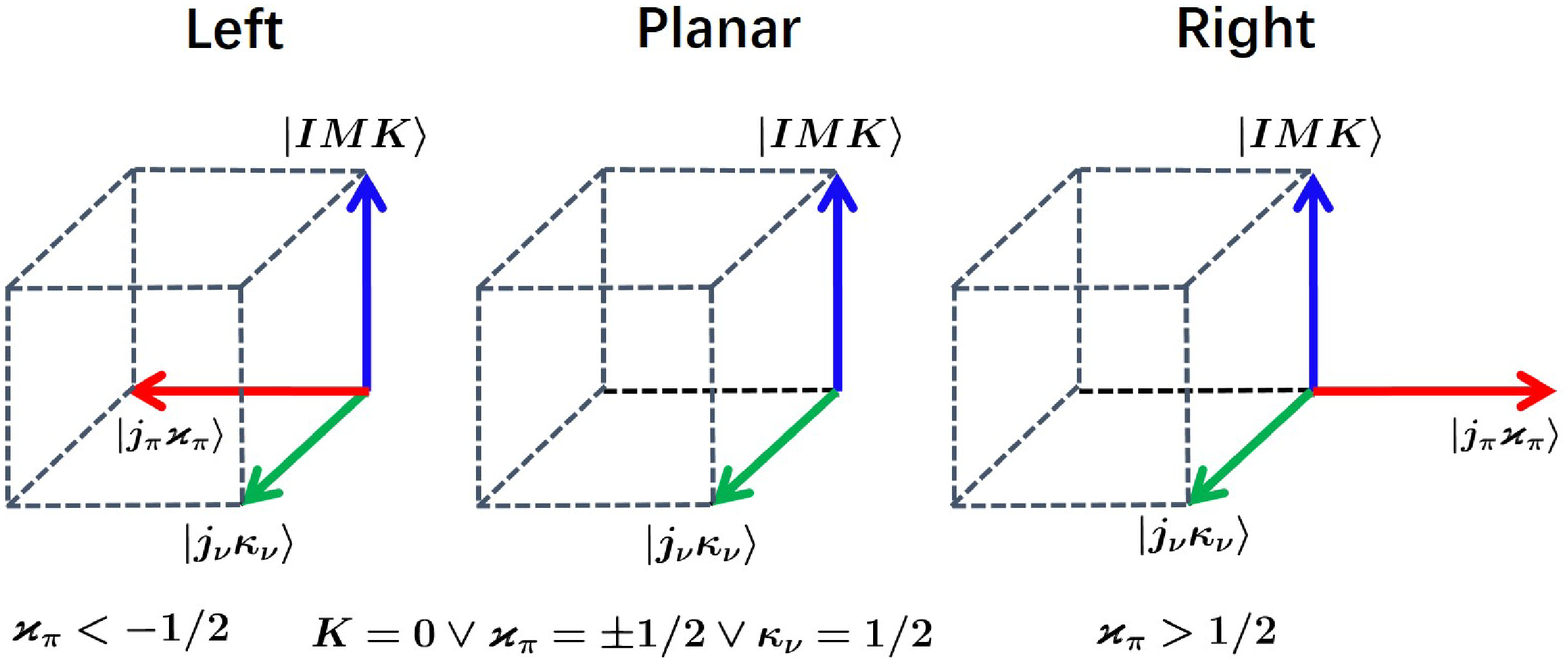}
    \caption{(Colour online) Cartoons of angular momentum coupling for
    the left-handed, planar, and right-handed bases.}\label{fig1}
  \end{center}
\end{figure}

 Among various nuclear models, the Particle Rotor Model
(PRM) has been widely used to describe the chiral doublet bands
achieving major successes~\cite{Frauendorf1997NPA, Starosta2001NPA,
  J.Peng2003PRC, Koike2004PRL, S.Q.Zhang2007PRC, B.Qi2009PLB,
  Lawrie2010PLB, Rohozinski2011EPJA, Shirinda2012EPJA}.  The model,
with the particular Hamiltonian for coupling of a proton particle
and a neutron hole considered here
\begin{equation}
H=H_R+H_\pi-H_\nu,
\end{equation}
describes in the laboratory reference frame a system consisting of
collective rotation of a triaxially deformed body
\begin{equation}
H_R=\sum_{i=1}^3\frac{R_i^2}{2\mathcal{J}_i}
\end{equation}
and single-particle motion of
valence nucleons in the body-fixed quadrupole-deformed mean field potential
\begin{eqnarray}
H_{\pi/\nu}&=&-\kappa(r)\beta\bigl[ \cos\gamma
Y_{2,0}(\theta,\phi)\bigr. \nonumber \\
&+&\bigl.\frac{1}{\sqrt{2}}\sin\gamma\left(Y_{2,2}(\theta,\phi)
+Y_{2,-2}(\theta,\phi)\right)\bigr]
\end{eqnarray}
with $\beta$ and $\gamma$ representing standard quadrupole-deformation
parameters and $\kappa(r)$ denoting radial dependence. The total
angular momentum of the system is
\begin{equation}
\vec{I}=\vec{R}+\vec{j}_\pi+\vec{j}_\nu.
\end{equation}

So far, chiral geometry of angular momentum coupling has been extracted
from expectation values of orientation operators, rather than being a
starting point at the outset of solving the PRM.  Very recently
Ref.~\cite{Starosta2017PS} proposed to solve the PRM in the Hilbert
space that contains subspaces of left-handed, right-handed, and planar
states of angular momentum coupling.
The basis states are defined by Eq.~95 of Ref.~\cite{Starosta2017PS}
\begin{eqnarray}
\label{Eq:tph}
&&\left| IMKj_\pi\varkappa_\pi j_\nu \kappa_\nu \right\rangle=\\
&=&\frac{1}{2}\sqrt{\frac{2I+1}{8\pi^2}}\left(D^I_{M K}(\upomega)
\left| j_\pi \varkappa_\pi \right\rangle
\left| j_\nu \kappa_\nu \right\rangle  \right.\nonumber\\
&+&(-1)^{I-j_\pi-\kappa_\nu} D^I_{M \overline{K}}(\upomega)
\left| j_\pi \overline{\varkappa_\pi} \right\rangle
\left| j_\nu \kappa_\nu \right\rangle\nonumber\\
&+&(-1)^{I+K+j_\nu-\varkappa_\pi+\kappa_\nu} D^I_{M \overline{K}} (\upomega)
\left| j_\pi \varkappa_\pi \right\rangle
\left| j_\nu \overline{\kappa}_\nu \right\rangle\nonumber\\
&+&\left.(-1)^{K+j_\pi+j_\nu -\varkappa_\pi} D^I_{M K} (\upomega)
\left| j_\pi \overline{\varkappa}_\pi \right\rangle
\left| j_\nu \overline{\kappa}_\nu \right\rangle\right),\nonumber\\
&&\overline{K}=-K,\;\; \overline{\varkappa_\pi}=-\varkappa_\pi,\;\;\overline{\kappa_\nu}=-\kappa_\nu,\nonumber\\
&&\mbox{for}\;K >0,\;
\varkappa_\pi\in[-j_\pi,j_\pi],
\;\kappa_\nu\in[\nicefrac{1}{2},j_\nu],\nonumber\\
&&\mbox{for}\;K =0,\;
\varkappa_\pi\in[\nicefrac{1}{2},j_\pi],
\;\kappa_\nu\in[\nicefrac{1}{2},j_\nu].\nonumber
\end{eqnarray}
with $K$, $\varkappa_\pi$ and $\kappa_\nu$ representing the projection
of $I$, $j_\pi$ and $j_\nu$ on the intermediate-, short-, and the
long-axis of the triaxial mean field, respectively.  The restrictions
on ranges of quantum numbers $K$, $\varkappa_\pi$ and $\kappa_\nu$ are
set to span the full Hilbert space while eliminating states which
differ by a phase only.  The choice of $K\geq 0$ and $\kappa_\nu \geq
\nicefrac{1}{2}$ is a convention adopted here following
Ref.~\cite{Starosta2017PS}.  For the above basis states, the
handedness is an explicit property defined by the sign of
$\varkappa_\pi$ as shown in Fig.~\ref{fig1} and is defined prior to
diagonalization of the Hamiltonian rather than extracted feature as in
models with non-chiral bases.

In details, the left-handed $\mathcal{L}$, the
right-handed $\mathcal{R}$, and the planar $\mathcal{P}$ subspaces
are defined as
\begin{eqnarray}
 \mathcal{L}&=&\{ K> 0 \wedge \varkappa_\pi <-1/2 \wedge \kappa_\nu >1/2\},\nonumber\\
\mathcal{R}&=&\{ K> 0 \wedge \varkappa_\pi > 1/2 \wedge \kappa_\nu > 1/2\},\nonumber \\
\mathcal{P}&=&\{ K=0 \vee
 \varkappa_\pi=\pm 1/2 \vee \kappa_\nu= 1/2\}.
\label{Eq:base}
\end{eqnarray}
The sum of the three subspaces is referred to as the ``Full'' Hilbert space.
Under the action of the operator $\mathcal{R}_2(\pi)\mathcal{T}$
the $\mathcal{L}$ and
$\mathcal{R}$ subspaces are transformed to each other, while the
$\mathcal{P}$ subspace is invariant.  The leading-order
Coriolis matrix elements are contained within each subspace
leading to nearly block-diagonal structure of the Hamiltonian matrix.
Current work explores impact of the coupling between
the planar and the chiral subspaces.

\begin{figure}[t]
  \begin{center}
    \includegraphics[width=5.5 cm]{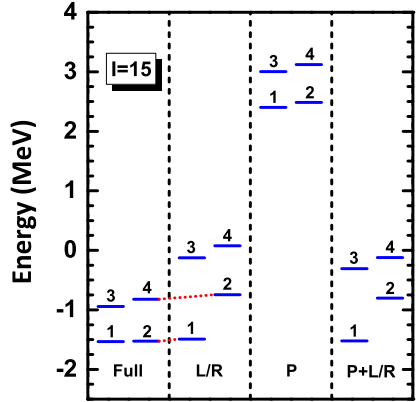}
    \caption{(Colour online) Energies from diagonalization of the PRM
      Hamiltonian at spin $I =15$ in combination of subspaces of
      Eq.\ \ref{Eq:base}.  The four combinations are: full basis
      (Full); left/right block ($\mathcal{L}/\mathcal{R}$), planar
      block ($\mathcal{P}$), and planar plus left/right blocks
      ($\mathcal{P}+\mathcal{L}/\mathcal{R}$). The four lowest energy
      states labelled as 1-4, respectively, are shown in each spectrum.
    }\label{fig2}
  \end{center}
\end{figure}

Figure\ \ref{fig2} presents results of block diagonalization of the
PRM Hamiltonian in the combinations of subspaces of Eq.\
\ref{Eq:base} in comparison to the diagonalization in the Full
basis.  The results are obtained for a symmetric particle-hole
configuration $\pi(1h_{11/2})^1\otimes\nu(1h_{11/2})^{-1}$
corresponding to the $A\approx 130$ mass region and the deformation
parameters $\beta=0.18$ and $\gamma=90^\circ$. The coupling constant
for the nuclear mean field potential defined by Eq. 13 of Ref.\
\cite{Starosta2017PS} has been assumed as 0.24~MeV/$\hbar^2$ while
the irrotational flow moments of inertia
$\mathcal{J}_3=4\mathcal{J}_1=4\mathcal{J}_2=\mathcal{J}_0$ are
adopted with $\mathcal{J}_0 = 30~\hbar^2/\textrm{MeV}$.  These
parameters result in formation of the chiral doublet bands and yield
degeneracy of states at the total spin value of $I=15^+$ as reported
previously in Refs.\ \cite{Starosta2017PS} and\
\cite{Frauendorf1997NPA}.  As seen in Fig.\ \ref{fig2}, the energies
resulting from the diagonalization in the Full space are the lowest,
which is expected for the complete basis.  The nearly degenerate
levels 1 and 2 as well as levels 3 and 4 are attributed to the
exactly degenerate solutions in the $\mathcal{L}$ and $\mathcal{R}$
subspaces. The lowest-energy solutions in the diagonalization in the
$\mathcal{P}$ subspace are separated from the lowest-energy
solutions in the $\mathcal{L}/\mathcal{R}$ subspaces by a sizable
$\approx 5~\textrm{MeV}$ energy gap. Consequently, the lowest-energy
state from block diagonalization in the
$\mathcal{P}+\mathcal{L}/\mathcal{R}$ space is nearly at the same
energy as levels 1 and 2 from the diagonalization in the Full space,
in good analogy to the DSPW model.

\begin{figure}[t]
  \begin{center}
    \includegraphics[width=7.5 cm]{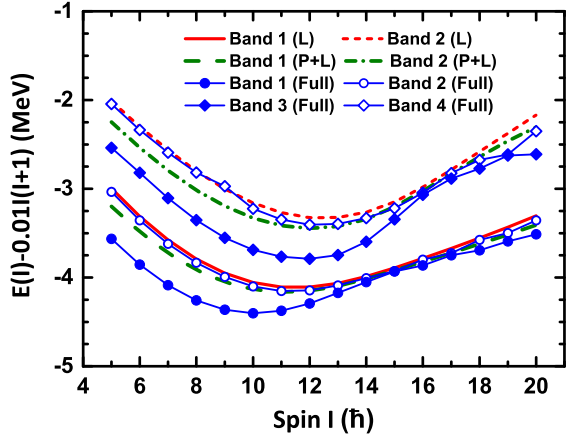}
    \caption{(Colour online) For the lowest two pairs of chiral doublet
      (bands 1-2 and 3-4) a comparison of diagonalization of
      the PRM Hamiltonian in the Full, $\mathcal{L}$, and
      $\mathcal{P}+\mathcal{L}$ subspaces.
    }\label{fig3}
  \end{center}
\end{figure}

In Fig.~\ref{fig2}, it is further observed that the levels 1 and 2
from diagonalization in the $\mathcal{L}/\mathcal{R}$ subspace are
nearly at the same energy as levels 2 and 4 from diagonalization in
the Full space, respectively.  The energy difference between the
states of the same spins in the doublet band are less than 200 keV in
a large spin window around the most degenerate levels at spin
15. Hence, the 5 MeV separation energy between the lowest-energy
solutions in the $\mathcal{L}$/$\mathcal{R}$ and $\mathcal{P}$
subspaces indicated by Fig.\ \ref{fig2} is indeed a sizable gap
compared with the energy of separation of the doublet band members.
Since calculations in Fig.\ \ref{fig2} are for spin $I=15$, a question
arises if this observation is representative for the whole spin
region. To investigate further, the corresponding lowest energy
spectra as a function of spin are shown in Fig.~\ref{fig3} with bands
labelled following the convention used in Fig.\ \ref{fig2}. It can be
observed in Fig.\ \ref{fig3} that bands 1 and 2 from diagonalization
in the $\mathcal{L}/\mathcal{R}$ subspace can very well reproduce
bands 2 and 4 from diagonalization in the Full space, respectively,
and that bands 1 and 2 from diagonalization in the
$\mathcal{P}+\mathcal{L}/\mathcal{R}$ subspace are lower in energy
than those from the diagonalization in the $\mathcal{L}/\mathcal{R}$
subspace.

To understand the above observation we propose a three energy level
model (3-ELM) for the Full diagonalization based on a simplification
which assumes that the $|\mathcal{L}\rangle$, $|\mathcal{R}\rangle$,
and $|\mathcal{P}\rangle$ subspaces consist of a single basis state
each. Note that for $\gamma=90^\circ$ $\langle
\mathcal{L}|H|\mathcal{R}\rangle =\langle
\mathcal{R}|H|\mathcal{L}\rangle=0$ and consequently the Hamiltonian
matrix takes the form
\small
\begin{align}
H
&=\left(
  \begin{array}{ccc}
    \langle \mathcal{L} |H|\mathcal{L}\rangle & \langle \mathcal{L}|H|\mathcal{R}\rangle
     & \langle \mathcal{L} |H|\mathcal{P}\rangle \\
    \langle \mathcal{R}|H|\mathcal{L}\rangle
     & \langle \mathcal{R} |H|\mathcal{R}\rangle & \langle \mathcal{R} |H|\mathcal{P}\rangle\\
   \langle \mathcal{P} |H|\mathcal{L}\rangle & \langle \mathcal{P} |H|\mathcal{R}\rangle
   & \langle \mathcal{P} |H|\mathcal{P}\rangle \\
  \end{array}
\right)
=\left(
  \begin{array}{ccc}
    \mathcal{A} & 0 & \mathcal{C} \\
    0 & \mathcal{A} & \mathcal{C}\\
   \mathcal{C}^* & \mathcal{C}^* & \mathcal{B} \\
  \end{array}
\right).
\label{Ham}
\end{align}
\normalsize
Results presented in Figs.~\ref{fig2} and \ref{fig3}
indicate $\mathcal{B} >>\mathcal{A}$. The energy-ordered
eigenvalues of Eq.\ \ref{Ham}
\small
\begin{align}
\lambda_1 &=\frac{1}{2}\Big[(\mathcal{B}+\mathcal{A})
 -\sqrt{(\mathcal{B}-\mathcal{A})^2+8\mathcal{C}^2}\Big],\notag\\
\lambda_2 &=\mathcal{A},\notag\\
\lambda_3 &=\frac{1}{2}\Big[(\mathcal{B}+\mathcal{A})
 +\sqrt{(\mathcal{B}-\mathcal{A})^2+8\mathcal{C}^2}\Big],
\label{eq1}
\end{align}
\normalsize
correspond to three eigenstates
\small
\begin{align}
\psi_1&=\frac{1}{\sqrt{N}_1}\left(
  \begin{array}{ccc}
   \displaystyle \frac{1}{4}\Big[(\mathcal{A}-\mathcal{B})
  -\sqrt{(\mathcal{A}-\mathcal{B})^2+8\mathcal{C}^2}\Big]\\
  ~~\\
   \displaystyle \frac{1}{4}\Big[(\mathcal{A}-\mathcal{B})
  -\sqrt{(\mathcal{A}-\mathcal{B})^2+8\mathcal{C}^2}\Big]\\
  ~~\\
   \mathcal{C}^*
  \end{array}
\right),\notag \\
\psi_2&=\frac{1}{\sqrt{2}}\left(
  \begin{array}{ccc}
  1\\
  -1\\
   0
  \end{array}
\right),\notag\\
\psi_3&=\frac{1}{\sqrt{N}_3}\left(
  \begin{array}{ccc}
    \displaystyle \frac{1}{4}\Big[(\mathcal{A}-\mathcal{B})
  +\sqrt{(\mathcal{A}-\mathcal{B})^2+8\mathcal{C}^2}\Big]\\
  ~~\\
    \displaystyle \frac{1}{4}\Big[(\mathcal{A}-\mathcal{B})
  +\sqrt{(\mathcal{A}-\mathcal{B})^2+8\mathcal{C}^2}\Big]\\
  ~~\\
   \mathcal{C}^*
  \end{array}
\right),
\label{wf3}
\end{align}
\normalsize
with $N_i$ denoting normalization of $\psi_i$ for $i=1$ and 3.

Based on the above results, it is interesting to note that
$\lambda_2=\mathcal{A}=\langle \mathcal{L} |H|\mathcal{L}\rangle
=\langle \mathcal{R} |H|\mathcal{R}\rangle$. This explains why the
energy spectra obtained by diagonalization in the
$\mathcal{L}/\mathcal{R}$ subspace reproduce part of the results of
diagonalization in the Full space, as shown in Figs.~\ref{fig2} and
\ref{fig3}.  Also, note that in the $\psi_2$ eigenfunction
$|\mathcal{L}\rangle$ and $|\mathcal{R}\rangle$ contributions mix
equally with the opposite phase forming an anti-symmetric
combination which contains only pure chiral components and does not
include any $|\mathcal{P}\rangle$ component.  Next, consider
eigenstates $\psi_1$ and $\psi_3$.  For these states
$|\mathcal{L}\rangle$ and $|\mathcal{R}\rangle$ components
contribute equally with the same phase and also the planar component
is admixed; this makes them distinctively different from the
$\psi_2$ state.  Due to the fact that $\mathcal{B}>>\mathcal{A}$ the
$|\mathcal{P}\rangle$ component in the $\psi_1$ state is much
smaller than $|\mathcal{L}\rangle$ and $|\mathcal{R}\rangle$
components. Correspondingly, the $\psi_3$ state is at high energy
($\approx 5~\textrm{MeV}$ higher than $\psi_1$) and with dominating
$|\mathcal{P}\rangle$ component.  High excitation energy is the most
likely reason why this state has not yet been observed.

Similarly, we can also construct a two energy level model (2-ELM) for
$\mathcal{P}+\mathcal{L}$ block diagonalization as
\small
\begin{align}
H
=\left(
  \begin{array}{cc}
   \langle \mathcal{L} |H|\mathcal{L}\rangle & \langle \mathcal{L} |H|\mathcal{P}\rangle \\
   \langle \mathcal{P} |H|\mathcal{L}\rangle & \langle \mathcal{P} |H|\mathcal{P}\rangle \\
  \end{array}
\right)
=\left(
  \begin{array}{ccc}
   \mathcal{A} & \mathcal{C} \\
   \mathcal{C}^*& \mathcal{B} \\
  \end{array}
\right).
\end{align}
\normalsize
The energy-ordered eigenvalues are
\small
\begin{align}
 \Lambda_{\mp} &=\frac{1}{2}\Big[(\mathcal{B}+\mathcal{A})
  \mp \sqrt{(\mathcal{B}-\mathcal{A})^2+4\mathcal{C}^2}\Big].
 \label{eq2}
\end{align}
\normalsize
Note that $\mathcal{B}>>\mathcal{A}$ implies $\lambda_1 <$ $\Lambda_{-}$ $< \lambda_2$.
For that reason block diagonalization in the
$\mathcal{P}+\mathcal{L}$ subspace yields energies lower than
block diagonalization in the $\mathcal{L}/\mathcal{R}$ or
$\mathcal{P}$ subspaces but higher than the diagonalization in the
Full space, as indeed illustrated in Figs.~\ref{fig2} and \ref{fig3}.

\begin{figure}[t]
  \begin{center}
    \includegraphics[width=7.5 cm]{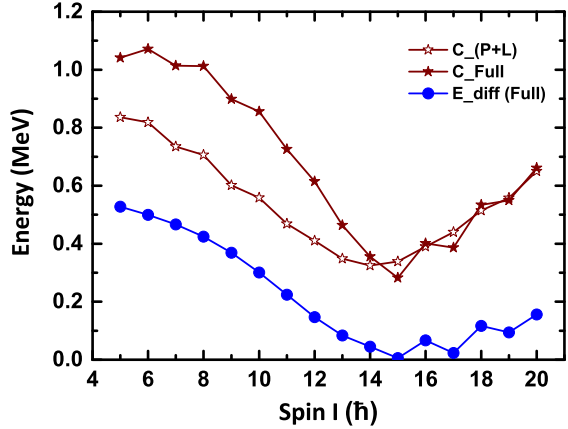}
    \caption{(Colour online) The effective interactions
      $\mathcal{C}=\langle\mathcal{L}|H| \mathcal{P}\rangle$ extracted
      from the three energy level model (3-ELM)
      ($\mathcal{C}_{\textrm{Full}}$) and the two energy level model
      (2-ELM) ($\mathcal{C}_{\mathcal{P}+\mathcal{L}}$) as a function
      of spin $I$, in comparison with the energy difference between
      the lowest pair of chiral doublet bands in the full
      diagonalization.}
\label{fig4}
  \end{center}
\end{figure}

Based on the conclusions drawn above, the values of effective
interaction $\mathcal{C}$ is extracted. In the 3-ELM according to
Eq.\ \ref{eq1}
\small
\begin{align}
 \mathcal{C}_{\textrm{Full}}=
\sqrt{\frac{1}{2}(\mathcal{A}-\lambda_1)(\mathcal{B}-\lambda_1)},
\label{c3}
\end{align}
\normalsize where $\mathcal{A}$, $\mathcal{B}$, and $\lambda_1$ are
taken as the lowest solution from diagonalization in the
$\mathcal{L}$, $\mathcal{P}$, and Full subspaces, respectively.
Similarly, one can extract $\mathcal{C}$ from the 2-ELM using
$\Lambda_{-}$ of Eq.\ \ref{eq2}. The results are presented in
Fig.~\ref{fig4} as a function of spin. The magnitude of the
interaction is $\mathcal{C} \approx 1$~MeV through the whole spin
region.

According to the 3-ELM if $\mathcal{C}=0$, then
$\lambda_2=\lambda_1$ and $\psi_1$ and $\psi_2$ are strictly
degenerate. In this case the $\psi_1$ and $\psi_2$ wave functions
contain only $|\mathcal{L}\rangle$ and $|\mathcal{R}\rangle$ chiral
components mixed with the same or the opposite phase but without any
$|\mathcal{P}\rangle$ contribution.  With the increase of
$\mathcal{C}$ the degeneracy is gradually broken. Consequently, the
magnitude of $\mathcal{C}$ determines the energy splitting between
the chiral doublet bands. Figure~\ref{fig4} presents the energy
differences between the lowest pair of chiral doublet bands as a
function of spin in comparison with the magnitude of $\mathcal{C}$
extracted using Eq.\ \ref{c3}.  The magnitude of $\mathcal{C}$ first
decreases and then increases as a function of spin. The turning
point is at $I\approx 15$ and $I\approx 14$ for
$\mathcal{C}_{\textrm{Full}}$ and
$\mathcal{C}_{\mathcal{P}+\mathcal{L}}$, respectively. The change of
the magnitude of $\mathcal{C}$ correlates very well with the
variation of energy difference between the members of the
lowest-energy doublet band. Particularly, for
$\mathcal{C}_{\textrm{Full}}$ an odd-even staggering behaviour is
extracted at the spin region $I\geq 15$ which is also observed for
the energy difference. In conclusion, the behaviour of the energy
difference between the doublet bands is determined by the
interaction matrix elements between the planar and non planar
subspaces of the full Hilbert space. To emphasize the analogy to the
DSPW model, we postulate that $1/\mathcal{C}$ corresponds to barrier
height between the left- and the right-handed potential wells. If
$\mathcal{C}$ approaches zero the barrier height approaches infinity
and chiral doublets become degenerate.  The increase of
$\mathcal{C}$ corresponds to the decrease of potential barrier and
thus leads to the interaction between $|\mathcal{L/R}\rangle$ and
$|\mathcal{P}\rangle$ subspaces which lifts the degeneracy between
doublet states.

\begin{figure}[t]
  \begin{center}
    \includegraphics[width=7.5 cm]{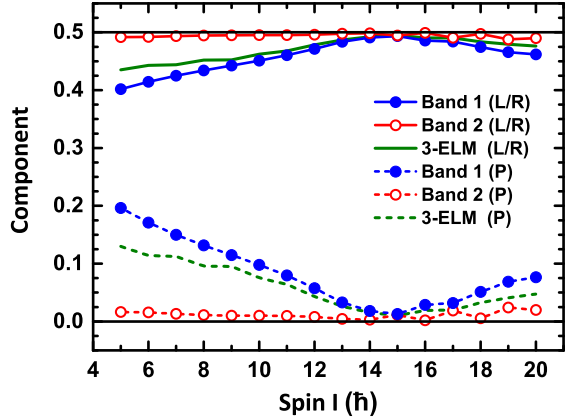}
    \caption{(Colour online) For the lowest energy chiral doublet band,
      a comparison of squared amplitudes of $\mathcal{L}/\mathcal{R}$
      and $\mathcal{P}$ components as a function of spin from
      the 3-ELM and from full diagonalization.}
    \label{fig5}
  \end{center}
\end{figure}

According to the 3-ELM states, band 2 do not contain the
$\mathcal{P}$ component, while the contribution of $\mathcal{P}$
component to states in band 1 is small relative to $\mathcal{L}$ and
$\mathcal{R}$ components. This is illustrated in Fig.~\ref{fig5}
showing a comparison of squared amplitudes of
$\mathcal{L}/\mathcal{R}$ and $\mathcal{P}$ components as a function
of spin for states in bands 1 and 2 from the 3-ELM and from full
diagonalization. It is observed that full diagonalization for band 2
leads to nearly vanishing contribution of $\mathcal{P}$ and
$\approx$50\% contribution of $\mathcal{L}$ and $\mathcal{R}$
components in full consistency with 3-ELM predictions. Consequently,
band 2 can be considered as of pure chiral geometry. For band 1 the
contribution of planar components shows a dependence on spin similar
to that for $\mathcal{C}$ in Fig.~\ref{fig4}, decreasing up to
$I=15$ and then increasing as a function of increasing spin.
Correspondingly, the $|\mathcal{L}\rangle$ or $|\mathcal{R}\rangle$
components increase and then decrease. At $I=15$ bands 1 and 2 are
separated by the smallest energy interval presenting the best case
of static chirality, nearly without the tunnelling between the $\mid
\mathcal{L} \rangle$ and $\mid \mathcal{R} \rangle$ components.
These analyses strongly suggest that chiral geometry is rather
robust as manifested in the band 2, while the degeneracy of states
with the same spin and parity occurs only in a limited spin range.
Again, properties of band 1 resulting from full diagonalization
agree well with predictions of the 3-ELM, which further indicates
that 3-ELM provides useful tools for understanding results of exact
diagonalization of the PRM.

In summary, three- and two-level mixing models are proposed to
understand the degeneracy of the chiral doublet bands.  It is
identified that the higher-energy member of a chiral doublet band is
formed by anti-symmetric combination of chiral components; it is of
pure chiral geometry without any planar components.  For the
lowest-energy band member planar components show first a decreasing
and then increasing trend with spin. This trend as well as the
magnitude of the energy splitting between both members are determined
by matrix elements of the Hamiltonian between the planar and nonplanar
subspaces of the full PRM Hilbert space.

The different planar components in the chiral doublet bands would be
used to interpret other possible different properties in the chiral
doublets, which provide impetus to reexamine the fingerprints of the
chiral doublets, such as electromagnetic transitions and angular
momentum geometries.

\begin{acknowledgements}
Q.B.C wants to thank F. Q. Chen, J. Meng, and S. Q. Zhang for
fruitful discussions. Financial support for this work was provided
in parts by the Major State 973 Program of China (Grant No.
2013CB834400), the National Natural Science Foundation of China
(NSFC) under Grants No. 11335002, No. 11375015, No. 11461141002, and
No. 11621131001, the China Postdoctoral Science Foundation under
Grants No. 2015M580007 and No. 2016T90007, the Deutsche
Forschungsgemeinschaft (DFG) and NSFC through funds provided to the
Sino-German CRC 110 ``Symmetries and the Emergence of Structure in
QCD'', and the Natural Sciences and Engineering Research Council of
Canada.
\end{acknowledgements}


\begin{thebibliography}{13}
\expandafter\ifx\csname
natexlab\endcsname\relax\def\natexlab#1{#1}\fi
\expandafter\ifx\csname bibnamefont\endcsname\relax
  \def\bibnamefont#1{#1}\fi
\expandafter\ifx\csname bibfnamefont\endcsname\relax
  \def\bibfnamefont#1{#1}\fi
\expandafter\ifx\csname citenamefont\endcsname\relax
  \def\citenamefont#1{#1}\fi
\expandafter\ifx\csname url\endcsname\relax
  \def\url#1{\texttt{#1}}\fi
\expandafter\ifx\csname urlprefix\endcsname\relax\def\urlprefix{URL
}\fi \providecommand{\bibinfo}[2]{#2}
\providecommand{\eprint}[2][]{\url{#2}}

\bibitem[{\citenamefont{Sakurai}(1994)}]{Sakurai1994book}
\bibinfo{author}{\bibfnamefont{J.~J.} \bibnamefont{Sakurai}},
  \emph{\bibinfo{title}{Modern Quantum Mechanics}}
  (\bibinfo{publisher}{Addison-Wesley Publishing Company},
  \bibinfo{year}{1994}).

\bibitem[{\citenamefont{Feyman}(1965)}]{Feyman1965book}
\bibinfo{author}{\bibfnamefont{R.~P.} \bibnamefont{Feyman}},
  \emph{\bibinfo{title}{The Feynman lectures on physics III: quantum
  mechanics}} (\bibinfo{publisher}{New York : Addison-Wesley},
  \bibinfo{year}{1965}).

\bibitem[{\citenamefont{Frauendorf and Meng}(1997)}]{Frauendorf1997NPA}
\bibinfo{author}{\bibfnamefont{S.}~\bibnamefont{Frauendorf}} \bibnamefont{and}
  \bibinfo{author}{\bibfnamefont{J.}~\bibnamefont{Meng}},
  \bibinfo{journal}{Nucl. Phys. A} \textbf{\bibinfo{volume}{617}},
  \bibinfo{pages}{131} (\bibinfo{year}{1997}).

\bibitem[{\citenamefont{Frauendorf}(2001)}]{Frauendorf2001RMP}
\bibinfo{author}{\bibfnamefont{S.}~\bibnamefont{Frauendorf}},
  \bibinfo{journal}{Rev. Mod. Phys.} \textbf{\bibinfo{volume}{73}},
  \bibinfo{pages}{463} (\bibinfo{year}{2001}).

\bibitem[{\citenamefont{Starosta et~al.}(2001)\citenamefont{Starosta, Koike,
  Chiara, Fossan, and LaFosse}}]{Starosta2001NPA}
\bibinfo{author}{\bibfnamefont{K.}~\bibnamefont{Starosta}},
  \bibinfo{author}{\bibfnamefont{T.}~\bibnamefont{Koike}},
  \bibinfo{author}{\bibfnamefont{C.~J.} \bibnamefont{Chiara}},
  \bibinfo{author}{\bibfnamefont{D.~B.} \bibnamefont{Fossan}},
  \bibnamefont{and} \bibinfo{author}{\bibfnamefont{D.~R.}
  \bibnamefont{LaFosse}}, \bibinfo{journal}{Nucl. Phys. A}
  \textbf{\bibinfo{volume}{682}}, \bibinfo{pages}{375c} (\bibinfo{year}{2001}).

\bibitem[{\citenamefont{Peng et~al.}(2003)\citenamefont{Peng, Meng, and
  Zhang}}]{J.Peng2003PRC}
\bibinfo{author}{\bibfnamefont{J.}~\bibnamefont{Peng}},
  \bibinfo{author}{\bibfnamefont{J.}~\bibnamefont{Meng}}, \bibnamefont{and}
  \bibinfo{author}{\bibfnamefont{S.~Q.} \bibnamefont{Zhang}},
  \bibinfo{journal}{Phys. Rev. C} \textbf{\bibinfo{volume}{68}},
  \bibinfo{pages}{044324} (\bibinfo{year}{2003}).

\bibitem[{\citenamefont{Koike et~al.}(2004)\citenamefont{Koike, Starosta, and
  Hamamoto}}]{Koike2004PRL}
\bibinfo{author}{\bibfnamefont{T.}~\bibnamefont{Koike}},
  \bibinfo{author}{\bibfnamefont{K.}~\bibnamefont{Starosta}}, \bibnamefont{and}
  \bibinfo{author}{\bibfnamefont{I.}~\bibnamefont{Hamamoto}},
  \bibinfo{journal}{Phys. Rev. Lett.} \textbf{\bibinfo{volume}{93}},
  \bibinfo{pages}{172502} (\bibinfo{year}{2004}).

\bibitem[{\citenamefont{Zhang et~al.}(2007)\citenamefont{Zhang, Qi, Wang, and
  Meng}}]{S.Q.Zhang2007PRC}
\bibinfo{author}{\bibfnamefont{S.~Q.} \bibnamefont{Zhang}},
  \bibinfo{author}{\bibfnamefont{B.}~\bibnamefont{Qi}},
  \bibinfo{author}{\bibfnamefont{S.~Y.} \bibnamefont{Wang}}, \bibnamefont{and}
  \bibinfo{author}{\bibfnamefont{J.}~\bibnamefont{Meng}},
  \bibinfo{journal}{Phys. Rev. C} \textbf{\bibinfo{volume}{75}},
  \bibinfo{pages}{044307} (\bibinfo{year}{2007}).

\bibitem[{\citenamefont{Qi et~al.}(2009)\citenamefont{Qi, Zhang, Meng, Wang,
  and Frauendorf}}]{B.Qi2009PLB}
\bibinfo{author}{\bibfnamefont{B.}~\bibnamefont{Qi}},
  \bibinfo{author}{\bibfnamefont{S.~Q.} \bibnamefont{Zhang}},
  \bibinfo{author}{\bibfnamefont{J.}~\bibnamefont{Meng}},
  \bibinfo{author}{\bibfnamefont{S.~Y.} \bibnamefont{Wang}}, \bibnamefont{and}
  \bibinfo{author}{\bibfnamefont{S.}~\bibnamefont{Frauendorf}},
  \bibinfo{journal}{Phys. Lett. B} \textbf{\bibinfo{volume}{675}},
  \bibinfo{pages}{175} (\bibinfo{year}{2009}).

\bibitem[{\citenamefont{Lawrie and Shirinda}(2010)}]{Lawrie2010PLB}
\bibinfo{author}{\bibfnamefont{E.~A.} \bibnamefont{Lawrie}} \bibnamefont{and}
  \bibinfo{author}{\bibfnamefont{O.}~\bibnamefont{Shirinda}},
  \bibinfo{journal}{Phys. Lett. B} \textbf{\bibinfo{volume}{689}},
  \bibinfo{pages}{66} (\bibinfo{year}{2010}).

\bibitem[{\citenamefont{Rohozinski et~al.}(2011)\citenamefont{Rohozinski,
  Prochniak, Starosta, and Droste}}]{Rohozinski2011EPJA}
\bibinfo{author}{\bibfnamefont{S.~G.} \bibnamefont{Rohozinski}},
  \bibinfo{author}{\bibfnamefont{L.}~\bibnamefont{Prochniak}},
  \bibinfo{author}{\bibfnamefont{K.}~\bibnamefont{Starosta}}, \bibnamefont{and}
  \bibinfo{author}{\bibfnamefont{C.}~\bibnamefont{Droste}},
  \bibinfo{journal}{Eur. Phys. J. A} \textbf{\bibinfo{volume}{47}},
  \bibinfo{pages}{90} (\bibinfo{year}{2011}).

\bibitem[{\citenamefont{Shirinda and Lawrie}(2012)}]{Shirinda2012EPJA}
\bibinfo{author}{\bibfnamefont{O.}~\bibnamefont{Shirinda}} \bibnamefont{and}
  \bibinfo{author}{\bibfnamefont{E.~A.} \bibnamefont{Lawrie}},
  \bibinfo{journal}{Eur. Phys. J. A} \textbf{\bibinfo{volume}{48}},
  \bibinfo{pages}{118} (\bibinfo{year}{2012}).

\bibitem[{\citenamefont{Starosta and Koike}(2017)}]{Starosta2017PS}
\bibinfo{author}{\bibfnamefont{K.}~\bibnamefont{Starosta}} \bibnamefont{and}
  \bibinfo{author}{\bibfnamefont{T.}~\bibnamefont{Koike}},
  \bibinfo{journal}{Phys. Scr.} \textbf{\bibinfo{volume}{92}},
  \bibinfo{pages}{093002} (\bibinfo{year}{2017}).

\end{thebibliography}
\end{document}